# Failure behaviors and processing maps with failure domains for hot compression of a powder metallurgy Ni-based superalloy

Zonglin Chi [1], Shuai Ren [1], Jingbo Qiao [1], Jinglong Qu [2], Chengbin Yang [2], Zhuanye Xie [3], Wei Chen [3], Hua Zhang [1], Liang Jiang [1], Shuying Chen [1, *], Fanchao Meng [1, *]

[1] *Institute for Advanced Studies in Precision Materials, Yantai University, Yantai Shandong 264005, China*

[2] *Wrought Superalloy Products Division, Gaona Aero Material Co., Ltd., Beijing 100081, China*

[3] *Nanshan Forge Company, Shandong Nanshan Aluminum Co., Ltd., Yantai Shandong 265700, China*

**Abstract:**

Processing maps are key to guiding the thermo-mechanical processing (TMP) of superalloys. However, traditional processing maps are incapable of delimiting failure, which is an essential factor to be concerned about during the TMP of superalloys. Employing isothermal hot compression experiments and finite element analysis (FEA), the present study examined the failure behaviors of a powder metallurgy (P/M) Ni-based superalloy and constructed processing maps with failure domains based on the predicted failure threshold. The micromechanical Gurson-Tvergaard-Needleman (GTN) damage model was employed in the FEA to model the cavity-driven intergranular fracture of the superalloy. Deformation temperature and strain rate were considered in the range of 1050 ~ 1150 °C and 0.001 ~ 1 s$^{-1}$, respectively. The FEA results reveal that the maximum tensile stress locates at the outer budging surfaces of the samples, which causes failure initiation and subsequent propagation into longitudinal cracks, being consistent with the experiments. It is further demonstrated that the failure is strain-controlled and the critical failure strain remains nearly insensitive to the range of strain rates considered while increasing with the increase of temperature in a third-order polynomial. Finally, an optimized processing window for hot deformation of the superalloy is formulated to warrant good hot workability while avoiding flow instability and failure. The present study offers direct insights into the failure behaviors of P/M Ni-based superalloys and details a modeling strategy to delineate optimized parametric spaces for the TMP of superalloys.

---

[*] Corresponding authors: Dr. Fanchao Meng, Email: mengfanchao@ytu.edu.cn; Dr. Shuying Chen, Email: sychen@ytu.edu.cn. Tel: +86-535-6902637.

## 1  Introduction

Ni-based superalloys possess exceptional tensile, creep, and fatigue properties at high temperatures [1-3], and thus have been predominantly employed for hot sections of gas turbine engines. Presently, the demand for more advanced engines able to operate at higher temperatures with fewer emissions has posed more stringent requirements on the properties of the Ni-based superalloys. Polycrystalline Ni-based superalloys are mainly comprised of A1 $\gamma$ matrix and the strengthening $L1_2$ $\gamma'$ phase, and their superior high-temperature properties are attributed to the solid solution strengthening within the $\gamma$ matrix and the precipitation strengthening from the ordered $\gamma'$ phase. Thus, the high-temperature properties can be improved by increasing the volume fraction of $\gamma'$ phase, which can be achieved by increasing the concentrations of the precipitate forming elements. However, elemental partition to $\gamma'$ phase could generate topologically close-packed (TCP) phases during high-temperature exposure, which are detrimental to the properties of the superalloys. In this regard, thanks to the high rates of solidification of the molten powders, powder metallurgy (P/M) is a more preferred processing method than the traditional ingot metallurgy to reduce the potential segregation [1, 4, 5].

The typical P/M processes include powder preparation, hot isostatic pressing (HIP), hot extrusion (HEX), forging, and heat treatment [1]. During the HIP process, prior particle boundaries (PPBs) on the surface of powder arising from a network of carbides, oxides, and oxy-carbides and thermally-induced porosities (TIP) will be present in the densified compact, which cannot be fully disrupted by the subsequent processes [6]. The existence of voids in the Ni-based superalloy in the present study was also confirmed by examining its microstructure (see Section 3.1 for details). These defects will most likely induce deformation cracks in the P/M superalloy when reaching a threshold strain, leading to a narrow processing window. He et al. [7] studied the forging cracks of a P/M Ni-based superalloy in the state of HEX condition, and found the existence of TIP and MC type carbides at grain boundaries accounted for the failure initiation through examining the microstructure evolution at different strains. They also employed dichotomy theory to determine the critical failure strains under different TMP conditions and found that the failure strain increased with the increase of temperature and/or strain rate. Wang et al. [8] discovered that the longitudinal surface cracks along the loading axis of the uniaxial compression originated from the intergranular cavities. Based on a theoretical analysis of the individual cavity in combination with experimental results, Zhang et al. [9] examined the dependence of the failure strain of forging cracks of P/M Ni-based superalloys on strain rate and found competition between intergranular fracture mode that is preferred under lower strain rates and shear bands mode that is dominated under higher strain rates. The failure strain increased with the increase of strain rate in the intergranular fracture mode and vice versa for the shear band mode. However, most of the experimental and theoretical studies concentrated on failure mechanisms and the tendencies of failure with TMP conditions. A thorough examination of the failure process and accurate prediction of the failure threshold of P/M Ni-based superalloys were still largely lacking.

As addressed by previous studies, the failure of the P/M Ni-based superalloys under a relatively low strain rate is dominated by the collective behaviors of cavities. To this end, the continuum-based micromechanical GTN model with damage criterion could be employed to describe structural damages and predict the failure threshold based on void nucleation, growth, and coalescence during the deformation of the P/M alloys [10-13]. For example, Mao et al. [14] employed the GTN damage model in the FEA to study the effect of notch ratios on multiaxial creep behavior and damage of Inconel 783 via effective porosity, demonstrating good agreement between experimental and FEA results. Zhao et al. [15] used the GTN damage model to simulate the influence of MC carbides on fracture behaviors of an Inconel 718 sheet during thermally assisted deformation. The GTN model has been also widely applied or extended for modeling the ductile fracture of other metals such as steel [16, 17], Al [18-20], Mg [21], and under a broad range of stress states [22-24]. More about the GTN model and its modifications can be referred to from a recent review by Lee and Basaran [25]. However, relatively few studies have been reported to probe the failure behaviors and thresholds of P/M superalloys under different TMP parameters by employing FEA with the GTN damage model.

As the crack formation is very sensitive to the TMP parameters (i.e., strain, strain rate, and temperature) [7, 26], it would be of importance to examine the effects of processing parameters on the failure behaviors and establish a processing window that considers failure to guide the TMP of the superalloys. Traditionally, it is well established that the hot workability of an alloy could be evaluated by the processing maps based on the dynamic materials model (DMM) [27], which are comprised of the efficiency of power dissipation to delimit the region of good workability and flow instability to define the region where plastic instability will occur. A further extension to the processing maps by coupling analysis of microstructural evolution was reported in the literature to instruct processing parameters [28-30]. In light of the concept of processing maps, presumably, a failure map could be constructed to further delineate the dominant parametric spaces for the hot workability of the superalloys. However, to the best of the authors' knowledge, such a map has not been reported for P/M Ni-based superalloys under different isothermal hot compression conditions.

Therefore, combining isothermal hot compression experiments and FEA, the present study examined the failure process of a P/M Ni-based superalloy to unveil the failure behaviors and predict the failure threshold under different TMP parameters. Based on the failure data, processing maps with failure domains were delineated to define the processing windows for the superalloy that allow good workability and meanwhile avoid instability and failure. In Section 2, the details about the experimental and computational methods will be elaborated, followed by experimental and FEA results and discussion in Section 3 and conclusions in Section 4. The present study established an FEA model able to quantitatively predict the failure threshold of P/M alloys during hot compression and clarified the dependence of failure threshold on TMP processing parameters, which is reflected in the processing maps, thus being able to guide the production practice of hot deformation process.

## 2 Experimental and computational methods

### 2.1 Experimental material

The material of interest is a Ni-based superalloy, with the chemical composition listed in **Table 1**. The alloy was manufactured via a P/M process including HIP of powders followed by HEX. The powders were manufactured by Aubert & Duval via vacuum induction melting and Argon atomization. Before the HIP process, the powders with a size smaller than 63 μm were screened out and followed by blending. The HIP process was conducted at 1160 °C for 4 h under a pressure of 150 MPa within a stainless-steel container to prepare a coupon, which was followed by the HEX process by putting the coupon into another stainless-steel container and sealing to obtain the extruded bar. The HEX was processed with an extrusion ratio of 3.5:1 and a temperature of 1130 °C. Cylinder samples with a diameter of 8 mm and a height of 12 mm were cut from the extruded bar via wire cutting for hot compression experiments (see **Fig. 1a**). Note that the samples were cut out in the same concentric circle of the extruded bar to minimize the differences in their microstructures.

**Table 1**. Chemical composition (wt%) of the Ni-based superalloy

| Element | Co | Cr | Mo | W | Al | Ti | Ta | C | B | Zr | Hf | Si | Ni |
|---|---|---|---|---|---|---|---|---|---|---|---|---|---|
| wt% | 13.13 | 11.96 | 4.02 | 4.03 | 2.99 | 4.03 | 4.02 | 0.055 | 0.04 | 0.028 | 0.14 | 0.03 | Bal. |

### 2.2 Experimental parameters

The isothermal hot compression experiments were carried out using Gleeble 3500-GTC thermal-mechanical testing system. High-temperature lubricant and Ta and graphite sheet were applied at both ends of the sample to prevent adhesion to the indenter as well as to reduce the influence of friction on the stress field of the sample during compression. Temperatures in the range of 1050 ~ 1150 °C and strain rates of 0.001, 0.01, 0.1, and 1 s$^{-1}$ were considered. For each testing temperature, the heating rate was 5 °C/s, and the temperature was held for 5 min before compression. In addition, to ensure the reliability of the experimental data, the deformation temperature was automatically compensated by the Gleeble machine. The total true strain was -0.69. The true stress-strain curve was outputted directly from the Gleeble machine. When the desired strain was achieved, the sample was immediately water quenched to retain the deformed microstructure. Microstructures of the samples were observed by using a scanning electron microscope (SEM, Tescan/Vega) with secondary (SE) and back scattered electron (BSE) detectors.

## 2.3 Constitutive model and failure criteria in FEA

In the GTN model, the void volume fraction, $f$, is defined as the ratio of the voids' volume to the material's total volume. The yield condition $\Phi$ as a function of $f$ of the GTN model is defined as [10, 11, 13]

$$\Phi = \left(\frac{q}{\sigma_y}\right)^2 + 2q_1 f^* \cosh\left(-q_2 \frac{3p}{2\sigma_y}\right) - (1 + q_3 f^{*2}) = 0 \qquad (1)$$

where $q = \sqrt{\frac{3}{2} \mathbf{S}:\mathbf{S}}$ is the effective Mises stress with $\mathbf{S}$ defined as $\mathbf{S} = p\mathbf{I} + \boldsymbol{\sigma}$, i.e., the deviatoric part of the Cauchy stress tensor $\boldsymbol{\sigma}$, where $p$ is the hydrostatic pressure, defined as $p = -\frac{1}{3}\boldsymbol{\sigma}:\mathbf{I}$. $\sigma_y(\bar{\varepsilon}_m^{pl})$ is the yield stress of the fully dense matrix material as a function of the equivalent plastic strain $\bar{\varepsilon}_m^{pl}$ in the matrix. $q_1, q_2, q_3$ are material parameters reflecting the interactions of the voids, which improve the model compatibility with the experiment [11]. The function $f^*(f)$ models the rapid loss of stress carrying capacity that accompanies void coalescence, which is defined as a function of the void volume fraction $f$ [13]

$$f^* = \begin{cases} f & \text{if } f \leq f_c \\ f_c + \frac{\bar{f}_F - f_c}{f_F - f_c}(f - f_c) & \text{if } f_c < f < f_F \\ \bar{f}_F & \text{if } f \geq f_F \end{cases} \quad (2)$$

where $\bar{f}_F = \left(q_1 + \sqrt{q_1^2 - q_3}\right)/q_3$, $f_c$ is a critical value of the void volume fraction, $f_F$ is the value of the void volume fraction at which there is a complete loss of stress-carrying capacity in the material. $f_c$ and $f_F$ model the material failure when $f_c < f < f_F$ due to mechanisms of nucleation, growth, and coalescence of voids. When $f \geq f_F$, total failure at the material point occurs. In Abaqus/Explicit, an element is removed once all of its material points have failed.

The evolution of the equivalent plastic strain in the matrix material is obtained from the equivalent plastic work [11]

$$(1 - f)\sigma_y \dot{\varepsilon}_m^{pl} = \boldsymbol{\sigma} : \dot{\boldsymbol{\varepsilon}}^{pl} \quad (3)$$

where $\dot{\boldsymbol{\varepsilon}}^{pl}$ is the plastic flow that is assumed to be normal to the yield surface and defined as $\dot{\boldsymbol{\varepsilon}}^{pl} = \dot{\lambda}\, \partial \Phi / \partial \boldsymbol{\sigma}$.

In terms of void growth and nucleation, the total change in void volume fraction, $\dot{f}$, is defined as [13]

$$\dot{f} = \dot{f}_{growth} + \dot{f}_{nucleation} \quad (4)$$

where $\dot{f}_{growth}$ is the change from the growth of existing voids and $\dot{f}_{nucleation}$ is the change due to the nucleation of new voids, which are defined as [12]

$$\dot{f}_{growth} = (1 - f)\dot{\boldsymbol{\varepsilon}}^{pl} : \mathbf{I} \quad (5)$$

$$\dot{f}_{nucleation} = A \dot{\varepsilon}_m^{pl} \quad (6)$$

with $A$ defined as [12]

$$A = \frac{f_N}{s_N \sqrt{2\pi}} exp\left[-\frac{1}{2}\left(\frac{\dot{\varepsilon}_m^{pl} - \varepsilon_N}{s_N}\right)^2\right] \quad (7)$$

where $\varepsilon_N$ and $s_N$ are the mean value and standard deviation of the normal distribution of the nucleation strain, respectively. $f_N$ is the volume fraction of the nucleated voids.

**2.4  FEA modeling parameters**

Abaqus/Explicit [31] was employed to model the hot compression process, which has fully integrated the GTN model with a failure definition into the code. The hot compression experiments were isothermal, and thus heat transfer was not simulated. To enable direct comparison between experimental and FEA results, the geometry and dimensions of the FEA model were kept the same as those of the experiment. The FEA model is shown in **Fig. 1b**, which is comprised of a cylinder with diameter and height being 8 and 12 mm, respectively, and two rigid plates contacting the cylinder. Considering that friction between the sample and the die will affect the expansion level and circumferential stress, two different yet reasonable friction coefficients, being 0.3 and 1.0, were considered to look into the effects of friction. Element type of C3D8R was selected for the cylinder and analytical rigid was set for the rigid plates.

Material properties were listed in **Table 2**. The density of the material was measured via the drainage method. Young's modulus was computed from a tensile test of the superalloy at room temperature. Poisson's ratio was assumed to be 0.3. Linear isotropic elasticity was specified. Note that the above material properties were assumed to be constant for the range of temperature considered in this study as the variation of these parameters is usually not significant to temperature. The isotropic yielding and hardening were assumed for the plasticity model of the fully dense matrix material. True stress-strain curves tested from the experiments were directly integrated into the simulation. Parameters in the first 6 lines of the GTN model, **Table 2**, were referred to Yang et al.'s [32] study on an FGH95 Ni-based superalloy, and in the last three lines were obtained based on the response surface method (RSM) by minimizing the difference between numerical and experimental force-displacement curves on a standard specimen [33, 34]. Particularly, the values of $q_1$, $q_2$, and $q_3$ were chosen to be the same with most of the studies in the literature [33], which were found to be reasonable for different types of metals and alloys.

Table 2. Material properties and parameters for FEA simulations

| Properties | Parameter | Unit | Value |
|---|---|---|---|
| General | Density | kg/m$^3$ | 8026 |
|  | Friction coefficient | … | 0.3 or 1.0 |
| Elastic | Young's modulus | MPa | 148000 |
|  | Poison's ratio | … | 0.3 |
| Plastic | Yield stress *vs.* plastic strain | … | From experiments, see Section 3.1 |
| GTN model | $q_1$ | … | 1.5 |
|  | $q_2$ | … | 1.0 |
|  | $q_3$ | … | 2.25 |
|  | $\varepsilon_N$ | … | 0.3 |
|  | $s_N$ | … | 0.1 |
|  | $f_N$ | … | 0.04 |
|  | $f_0$ | … | 0.005 |
|  | $f_c$ | … | 0.02 |
|  | $f_F$ | … | 0.04 |

During the FEA simulation, the bottom rigid plate was held fixed and the top rigid plate was given a velocity corresponding to one of the four strain rates, including 0.001, 0.01, 0.1, and 1 s$^{-1}$. A displacement boundary condition was imposed on the top die to compress the sample to its half-height, which corresponded to a true strain of -0.69. Five temperatures of the cylinder were considered, including 1050, 1075, 1100, 1125, and 1150 °C. Therefore, in total 20 simulations were performed.

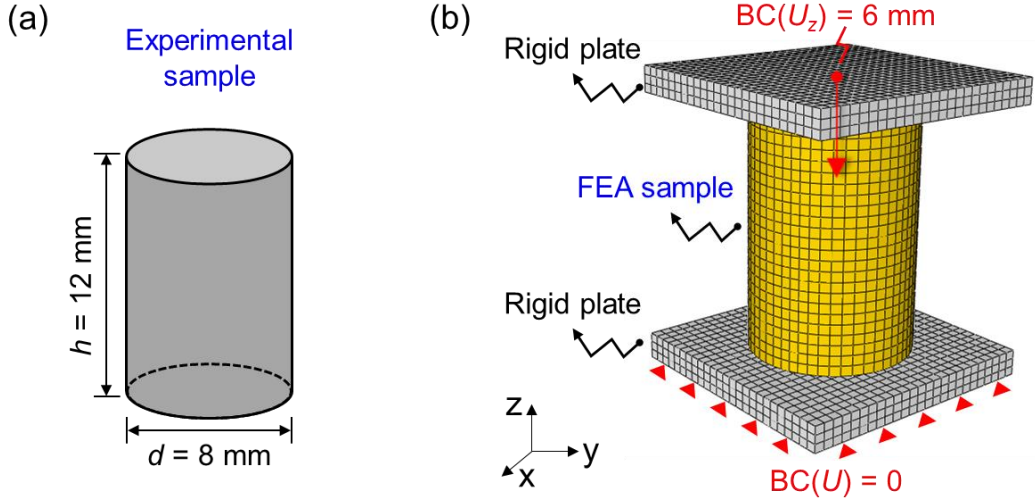

**Figure 1**. Schematics showing (a) the experimental sample with dimensions denoted and (b) the FEA model and its boundary conditions, where the sample has the same dimensions as the experiment and is contacting with two rigid plates.

## 2.5  Construction of the processing maps

Processing maps were constructed based on the DMM [27]. The model treats the hot working of a material to be a process of dissipation of power, which occurs through a temperature rise ($G$ content) and a microstructural change ($J$ co-content). The partition between the two dissipation is determined by a strain-rate sensitivity factor ($m$) of flow stress ($\sigma$). The strain-rate sensitivity of a material is evaluated as [27]

$$m = \frac{\partial \ln \sigma}{\partial \ln \dot{\varepsilon}} \tag{8}$$

where $\dot{\varepsilon}$ is the strain rate. At a given temperature and strain, the $J$ co-content is defined as [27]

$$J = (\sigma \times \dot{\varepsilon} \times m)/(m + 1) \tag{9}$$

A dimensionless parameter, $\eta$, normalized from the non-linear dissipator $J$ by an ideal linear dissipator ($m=1$) is given by [27]

$$\eta = J/J_{max} = 2m/(2m+1) \tag{10}$$

The parameter $\eta$ is called efficiency of power dissipation, the variation of which with temperature and strain rate represents the constitutive behavior of the material and constitutes a power dissipation map.

The flow instability during hot deformation was defined in the continuum criterion developed by Kumar and Prasad [27, 35] by combining the extremum principles of irreversible thermodynamics (the basis of the DMM) and the separability of power dissipation. The instability parameter, $\xi(\dot{\varepsilon})$, is defined in Eq. (11) [35] and the flow instability will occur if the parameter is small than 0.

$$\xi(\dot{\varepsilon}) = \frac{\partial \ln[m/(m+1)]}{\partial \ln \dot{\varepsilon}} + m < 0 \tag{11}$$

The variation of the instability parameter with temperature and strain rate constitutes the instability map. In addition, the failure strains numerically predicted from FEA simulations for hot compression of the Ni-based superalloy in the present study constitute the failure map, which will be superimposed on the power dissipation and instability maps to construct the processing maps with failure domains.

## 3    Results and Discussion

### 3.1    Hot compression experiment

Representative microstructures of the samples before compression are shown in **Fig. 2**, where it can be seen that inside the sample there exist some voids. Such defect is one of the most common defects in P/M superalloys, which was caused by insoluble gases, such as Ar, and expanded during the HIP thermal process.

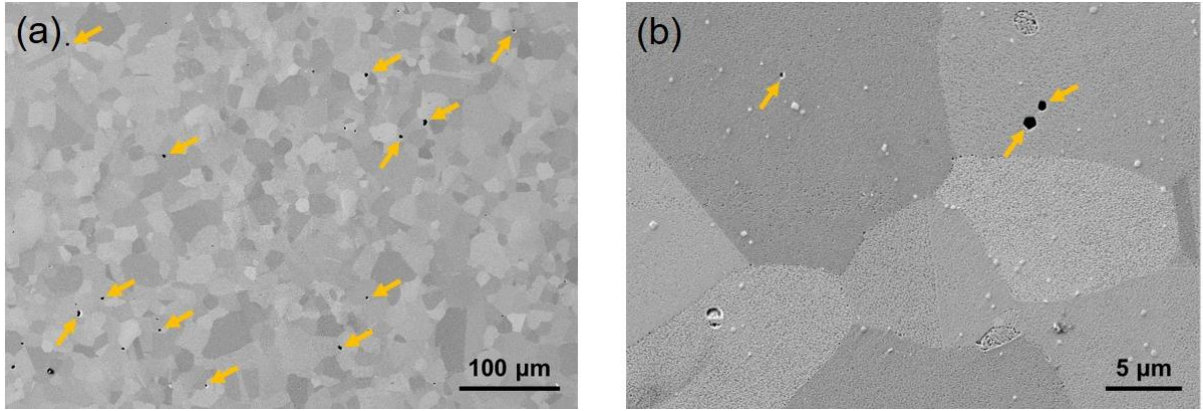

**Figure 2**. Microstructures of the Ni-based superalloy prior to hot compression under resolutions of (a) 100 μm and (b) 5 μm, respectively. The orange arrows point out fabrication-induced voids in the P/M superalloy.

**Fig. 3** plots the experimental true stress-strain curves of the samples under different deformation conditions. It can be observed that the flow stress first increases with strain to reach a peak stress, which is because work hardening arising from the increase and accumulation of dislocations dominates over the dynamic recovery (DRV) from dislocation motions. Past the peak stress, for strain rates of 0.001 and 0.01 s$^{-1}$ and temperatures of 1125 and 1150 °C, the flow stress plateaus with strain, due to the balance between work hardening and dynamic softening from DRV. While, for the other deformation conditions, the flow stress gradually decreases with the increase of strain. This phenomenon can be attributed to dynamic recrystallization (DRX), where the dislocation densities at the center of the recrystallized grains have increased sufficiently to promote another cycle of nucleation and grain growth till the recrystallization is complete, thus increasing dynamic softening. Then, with the continued increase of strain, the flow stress plateaus to a level intermediate between the peak and yield stresses (i.e., rheological stability phenomenon), originating from the balance between work hardening and dynamic softening from DRX. Overall, the stress-strain responses of the Ni-based superalloy exhibit characteristics of typical DRV and DRX rheological curves. In addition, it is worth noting that the stress-strain curves show serrated features, which is likely due to the emergence of dynamic strain aging (DSA).

Moreover, the strain corresponding to the peak stress in the flow curve can be treated as a sign of the onset of DRV/DRX, the occurrence of which modifies the appearance of the flow curve. The strain can be denoted as the peak strain. Comparing the peak strains, the lower the temperature, the larger the peak strain, and the higher the strain rate, the larger the peak strain, indicating that the DRV/DRX necessitates a large strain to be triggered when the deformation temperature is low and/or the strain rate is high.

Furthermore, the flow stress is decreasing with the increase in temperature. When the temperature increases from 1050 to 1150 °C, the magnitude of the peak stress drops ~30%. Moreover, the flow stress is also greatly dependent on strain rate. The lower the strain rate, the lower the stress. The peak stress at a strain rate of 0.001 $s^{-1}$ retains only ~10% of that at a strain rate of 1 $s^{-1}$. Although both temperature and strain rate greatly affect the deformation behaviors of the material, it is worth noting that the variations of the temperature alter the characteristics of the flow curves, while at the same temperature, the shapes of the stress-strain curves are similar for different strain rates.

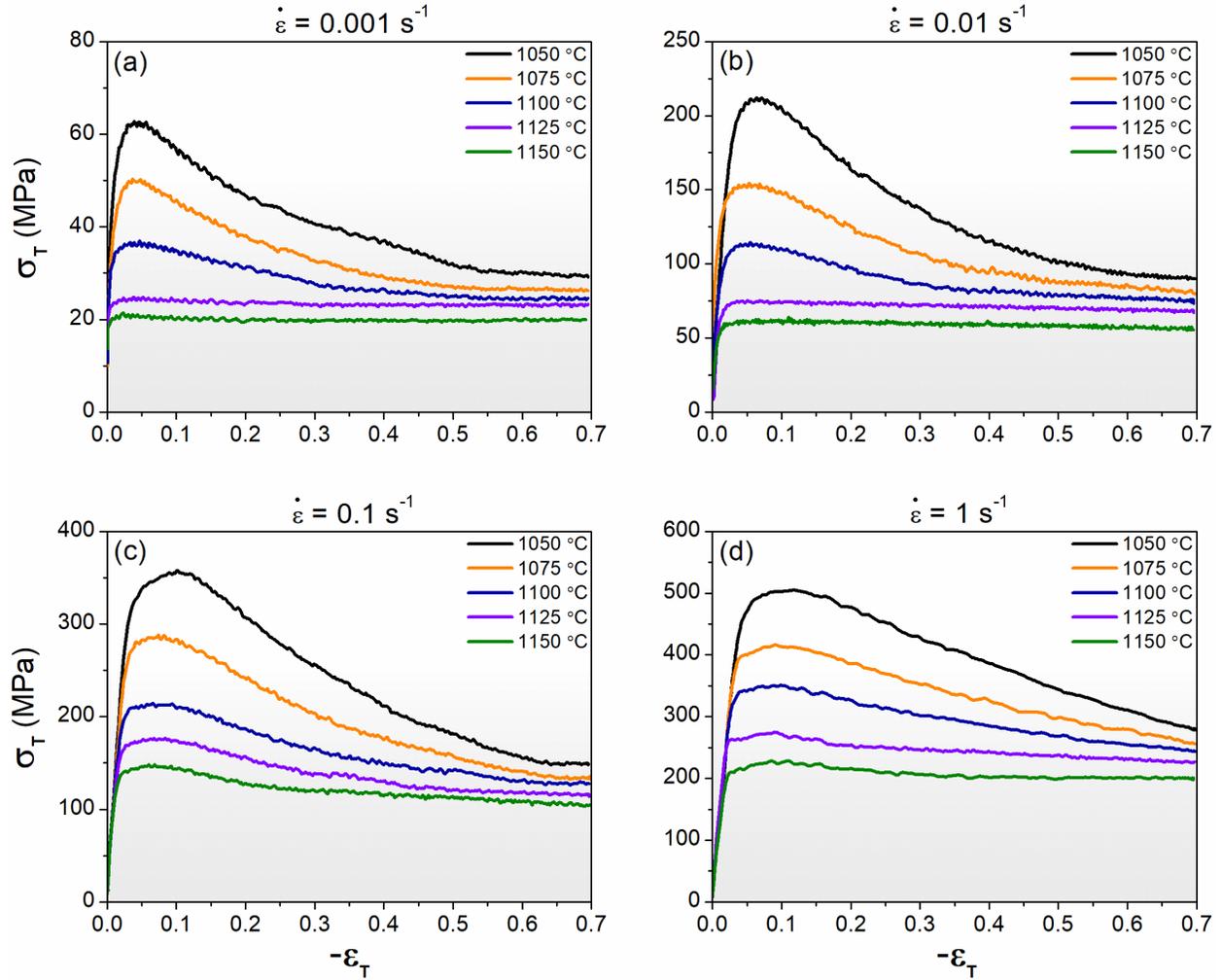

**Figure 3**. True stress-strain curves of the samples as a function of strain rate in (a) 0.001 s$^{-1}$, (b) 0.01 s$^{-1}$, (c) 0.1 s$^{-1}$, and (d) 1 s$^{-1}$ obtained during hot compression experiments. Five deformation temperatures in the range from 1050 to 1150 °C are considered.

The deformed configurations of the samples that show cracks are examined, as shown in **Fig. 4a**. It is seen that some samples under higher strain rates, especially at the strain rate of 1 s$^{-1}$, exhibit severe cracks around the outmost surface. In addition, consistent with the analysis of the true stress-strain curves, the samples experience more severe failure under lower temperatures and/or higher strain rates, which may be owing to the severe work hardening under such conditions. The representative microstructures of the center positions of the samples under 1100 °C and 1 s$^{-1}$ hot compression till true strains (denoted as $\varepsilon_T$ hereafter) of -0.4, -0.5, and -0.6 that were cut parallelly to the outmost bulged surface are shown in **Fig. 5**. Compared with the microstructures before deformation in **Fig. 2a-b**, the deformed microstructures exhibit the growth and coalescence of the initial voids into large flaws accompanying deformation, being consistent with He et al.'s [7] study. Particularly at $\varepsilon_T = -0.6$, there exist evident large pores/holes as well as small pores around the large pores, which are expected to cause the failure of the sample.

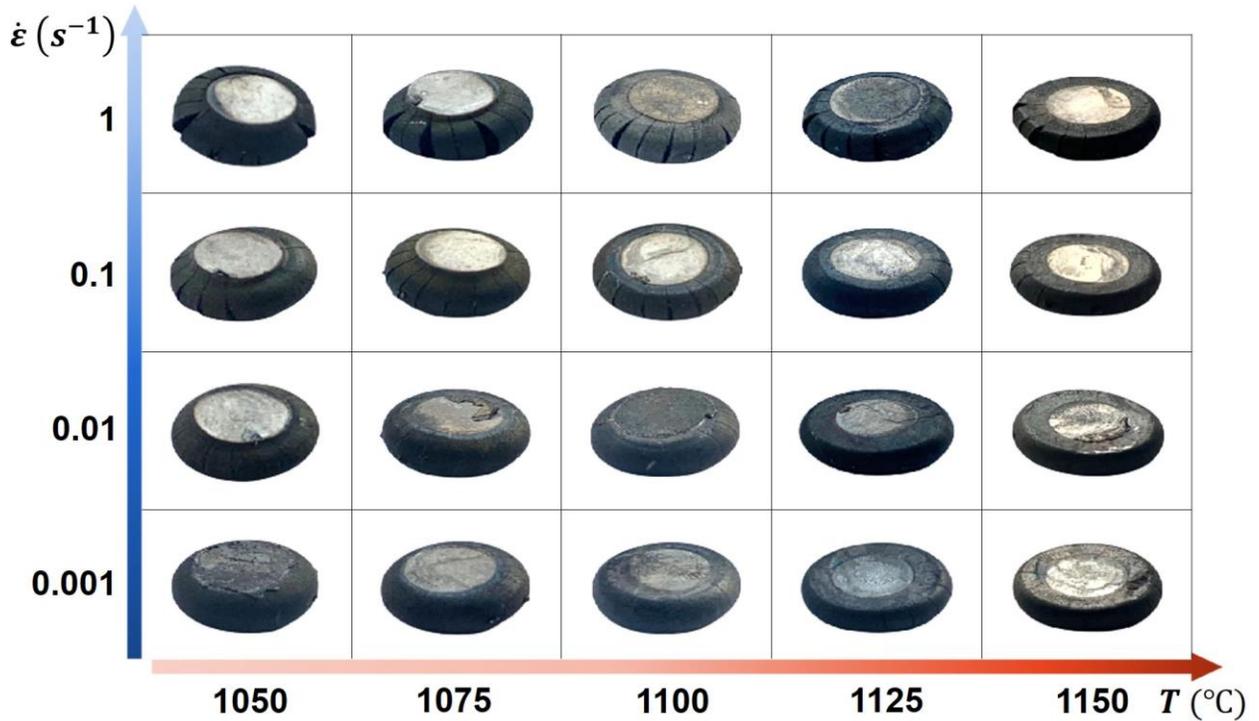

**Figure 4**. Deformed configurations of samples under different deformation conditions.

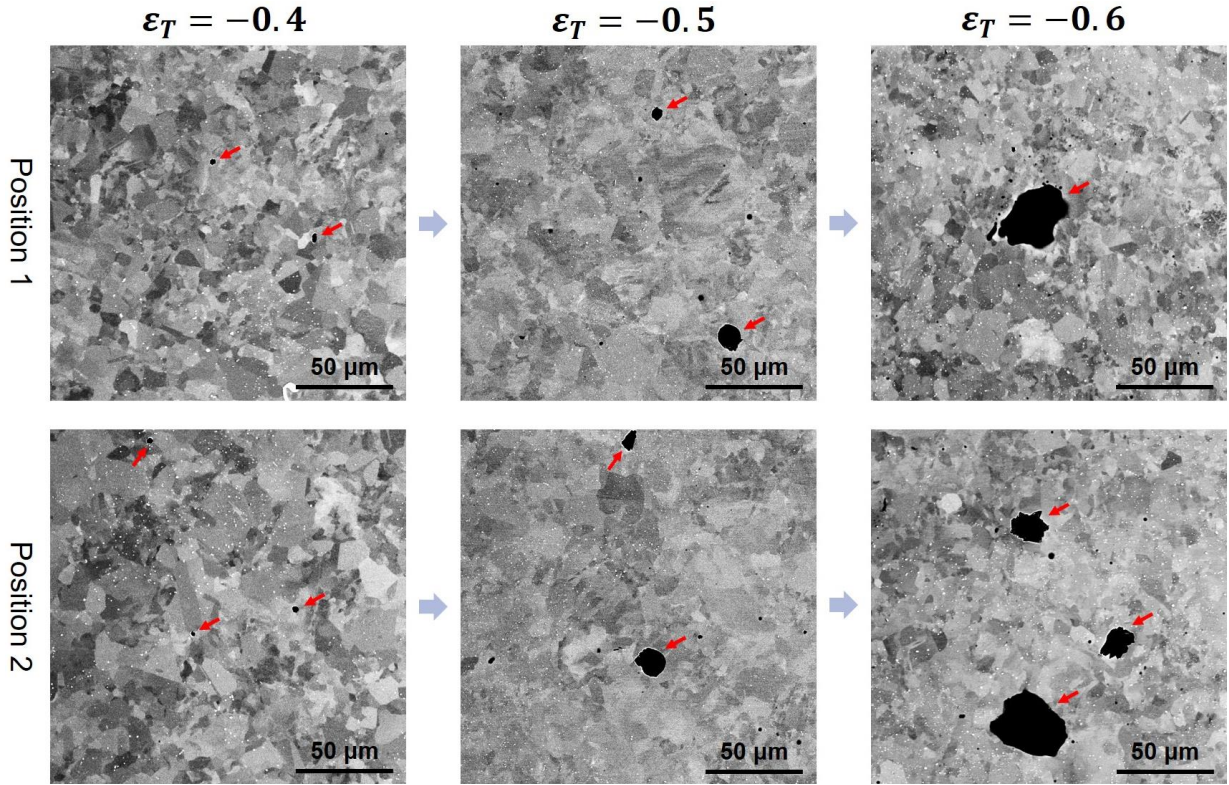

**Figure 5**. Microstructure evolution with respect to the true strain under deformation temperature of 1100 °C and strain rate of 1 s$^{-1}$. Representative positions 1 and 2 are selected as demonstration. The red arrows point out the pores/holes.

### 3.2  FEA simulated hot compression

The deformed configurations of the samples with the contour of the maximum principal stress simulated by FEA up to $\varepsilon_T$ = -0.36 under different hot compression conditions are displayed in **Fig. 6**. It can be observed that all the samples develop similar stress contours that are the maximal tensile stress concentrated in the mid-region of the outer bulging surface, as delimited by the dashed line box in **Fig. 6**, and the maximal compression stress located in the body center of the sample. It thus can be speculated that with the increase of the external compressive strain, the local tensile stress would firstly cause the failure of the material points at the outer bulging surface, which is consistent with the observations of the experiments.

In addition, consistent with the aforementioned analysis, the samples experience more severe tensile stress under lower temperatures due to more concentrated dislocations and higher strain rates due to shorter time for dislocation slips. The maximum principal stress is 330, 280, 240, 180, and 140 MPa for deformation temperatures increasing from 1050 to 1150 °C, respectively. The percentage of stress decrease is 57.6% at 1150 °C relative to that of 1050 °C. Nevertheless, it can hardly notice the difference among the stress contours of different strain rates. In this regard, temperature dominates the evolution of the local stress in the sample over the strain rate. Furthermore, it is noteworthy that the variation of friction coefficients between 0.3 and 1.0 does not noticeably affect the stress contours of the samples after hot compression. The influence of friction coefficients is quantitatively examined in **Fig. 11**, and for simplicity, only the results from the friction coefficient of 0.3 are exhibited in **Fig. 6**.

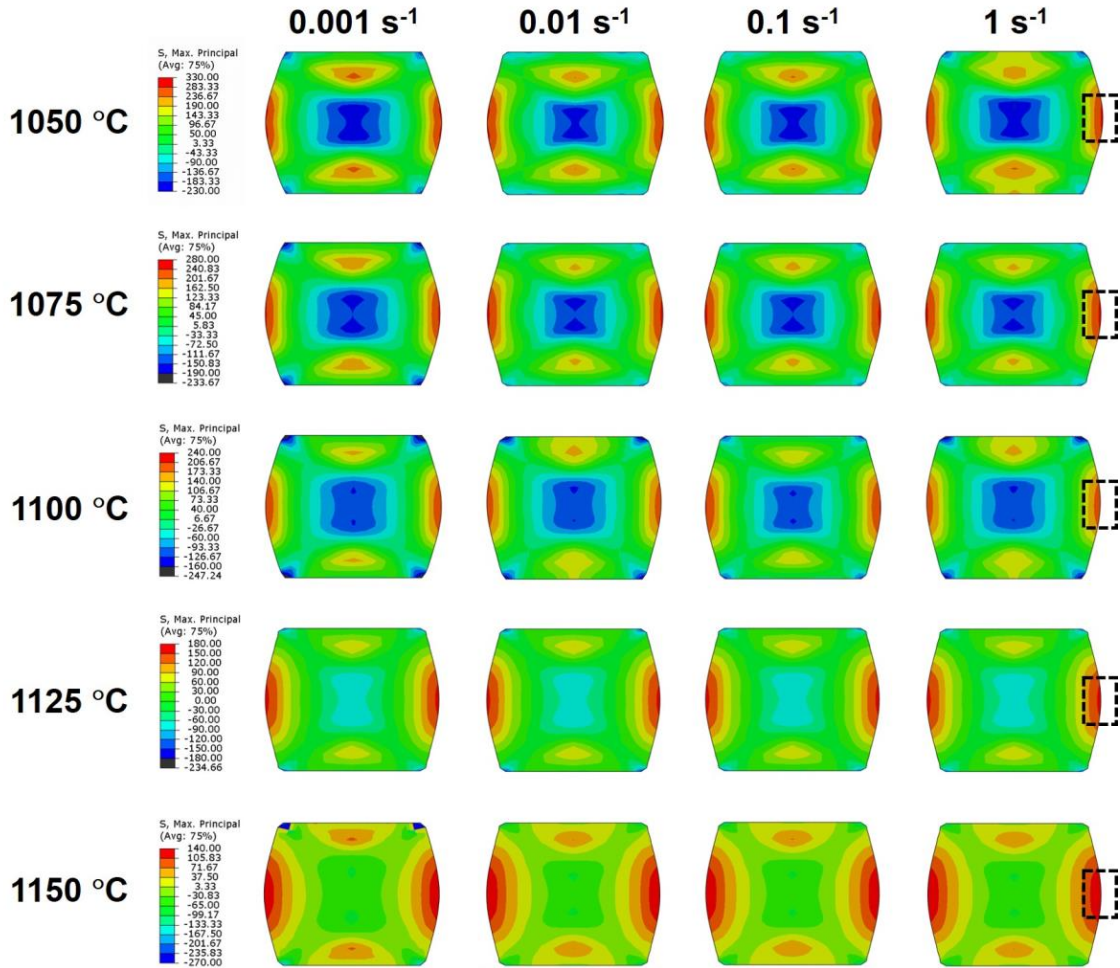

**Figure 6**. Deformed configurations of samples with contours of the maximum principal stress at a true strain of -0.36 under different hot compression conditions (with a friction coefficient being 0.3). The scale bar remains to be the same for the same deformation temperature. A dashed line box delimited the location of the largest maximum principal stress within the body of the sample.

To investigate the stress evolution within the dashed line box in **Fig. 6**, the maximum principal stress of the mid-point at the outer bulging surface of the sample as a function of true strain for different hot compression conditions is plotted in **Fig. 7**. It can be seen that the stress increases largely linearly with the increase of strain, and increases more rapidly for lower temperatures than for higher temperatures. In addition, the evolution of the maximum principal stress among different strain rates is nearly the same for the same deformation temperature. These observations may indicate that temperature is a more controlling processing factor than the strain rate in the failure behaviors of the P/M Ni-based superalloy.

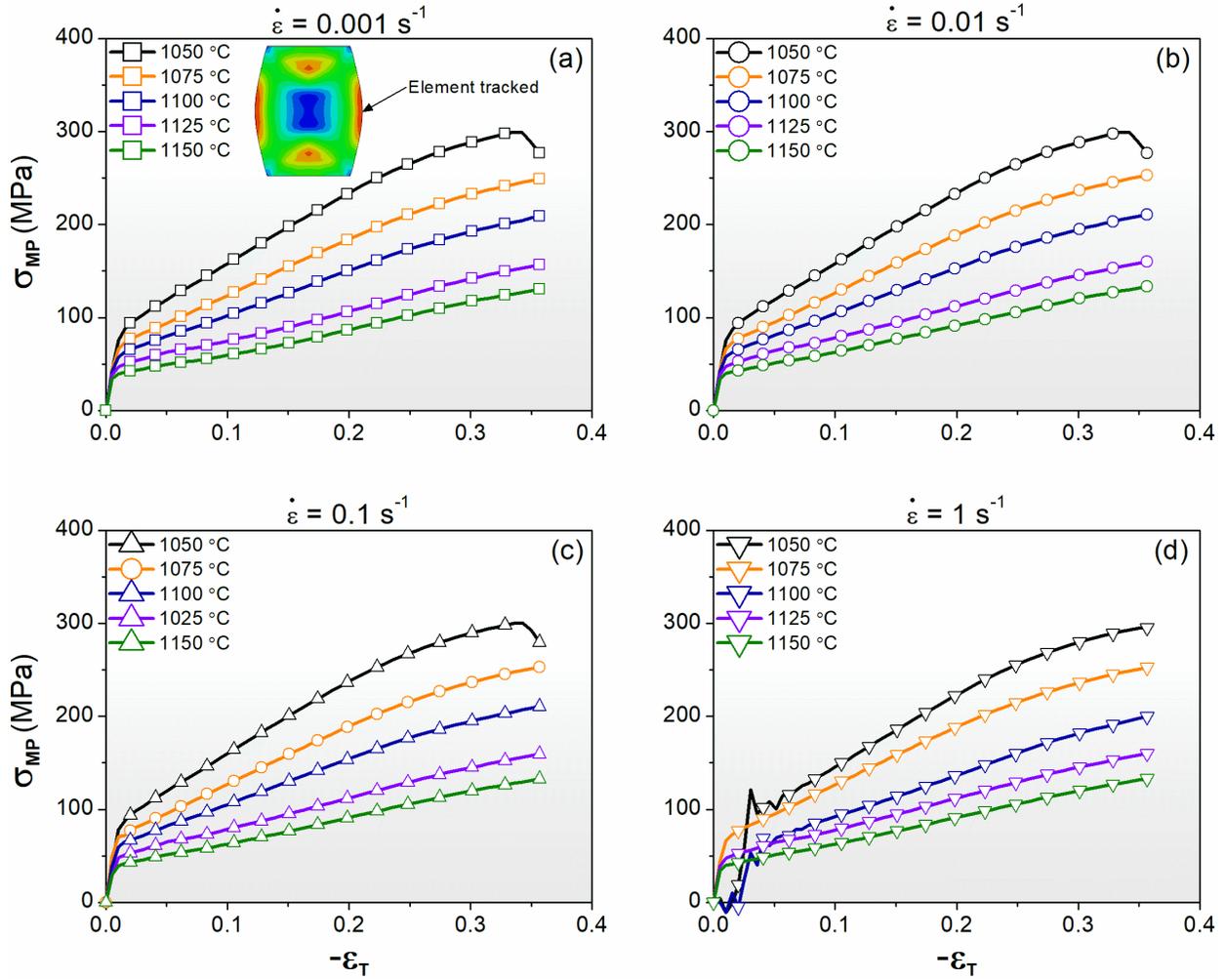

**Figure 7**. The evolution of the maximum principal stress of the mid-point of the outer bulging surface as a function of true strain up to a true strain of -0.36 in (a) 0.001 s$^{-1}$, (b) 0.01 s$^{-1}$, (c) 0.1 s$^{-1}$, and (d) 1 s$^{-1}$. The deformation temperatures are in the range of 1050 to 1150 °C and the friction coefficient is 0.3.

To further examine the deformation and failure process of the samples with the increase of external strain, four representative deformation conditions are selected to cover low temperature/strain rate to high temperature/strain rate. From **Fig. 8a-b**, it can be seen that when the magnitude of the true strain increases from 0 to 0.24, the mid-regions of the outer bulging surface develop strong and localized tensile maximum principal stress. This behavior can be ascribed to the fact that a larger strain will induce higher degrees of bulging that result in increased tensile stress at the outer surface of the sample. When $-\varepsilon_T$ increases to 0.41, a few elements at the surface of the samples undergo failure and are deleted from the body of the sample, followed by failure propagation and coalescence into long stripes along the loading direction when $-\varepsilon_T = 0.65$. Comparing **Fig. 8a** with **Fig. 8b**, the strain rate does not affect the stress contour of the sample. However, lower strain rate results in more failure than a higher strain rate at 1050 °C when $-\varepsilon_T = 0.41$. These observations are in line with the study of He et al. [7], where it was reported that there existed a critical strain for the onset of failure and a lower strain rate caused the earlier failure. The result is also consistent with Zhang et al.'s study [9], where the failure strain decreases with the decrease of strain rate under the intergranular fracture mode. The mechanism underlying such behavior will be elaborated in **Fig. 12**. The critical failure strain, denoted as $\varepsilon_f$, thus can be extracted from the FEA simulation by finding the strain that corresponds to the first failure of the element. Besides, $\varepsilon_f$ can also be determined based on the evolution of void volume fraction due to growth and due to nucleation (see **Figs. 9-10** for details).

Comparing **Fig. 8c-d** with **Fig. 8a-b**, it can be observed that the sample develops lower stress with the increase of external strain when the temperature is higher. This is an echo of what is discussed earlier that temperature dominantly determines the stress evolution of the sample.

Moreover, the sample remains intact when $-\varepsilon_T = 0.41$, and failure does not occur until $\varepsilon_T$ is around -0.65 for $T = 1150$ °C, albeit higher strain rate results in more failures than a lower strain rate, being opposite to the scenarios when $T = 1050$ °C. Nevertheless, the overall failure trend evolving with temperature and strain rate agrees well with that observed in experiments (see **Fig. 4**). Besides, the deformed configurations of the sample and the appearance of the stripe failure from the FEA simulations also highly agree with those of the experiments. Therefore, although it is difficult to directly measure the failure strain in the experiment to verify the FEA result, the present FEA simulation shall be able to quantitatively predict the failure threshold for hot deformation of the P/M superalloy.

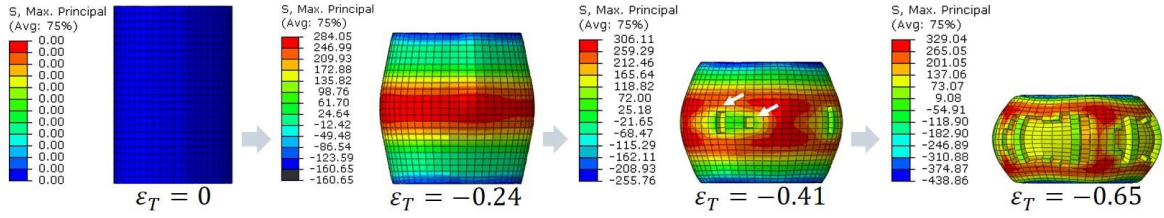

(a) T = 1050 °C, $\dot{\varepsilon} = 0.001\ s^{-1}$

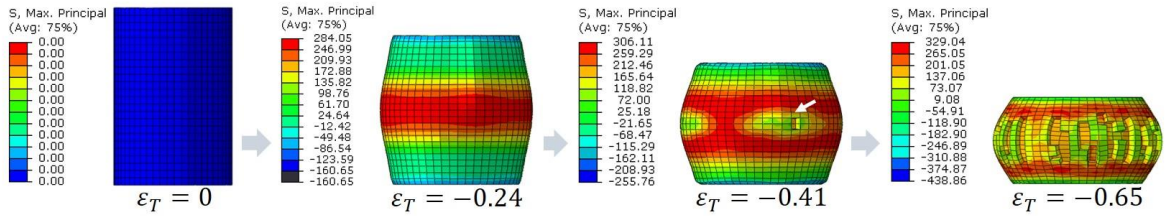

(b) T = 1050 °C, $\dot{\varepsilon} = 0.1\ s^{-1}$

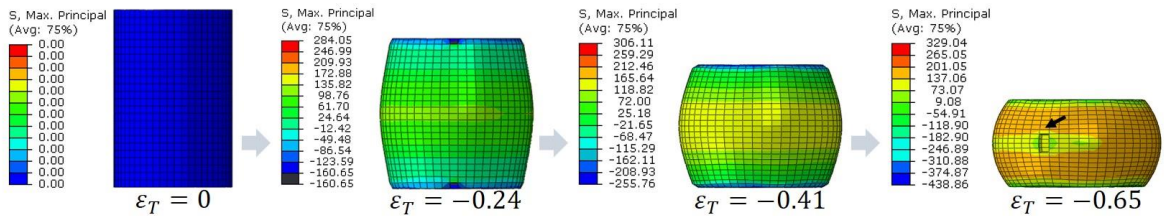

(c) T = 1150 °C, $\dot{\varepsilon} = 0.001\ s^{-1}$

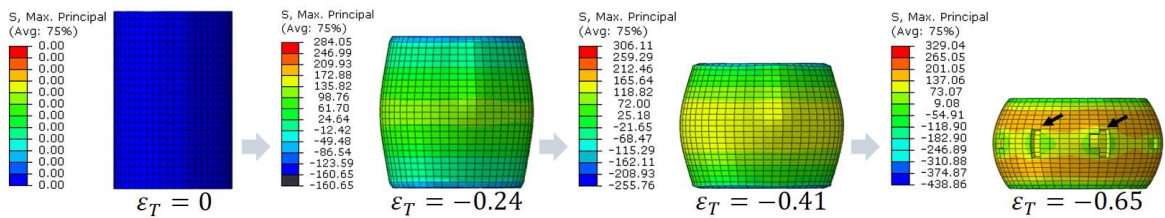

(d) T = 1150 °C, $\dot{\varepsilon} = 0.1\ s^{-1}$

**Figure 8**. The deformation and failure processes of four representative samples concerning the increase of true strain under different conditions in (a) $T$ = 1050 °C and $\dot{\varepsilon}$ = 0.001 s$^{-1}$, (b) $T$ = 1050 °C and $\dot{\varepsilon}$ = 0.1 s$^{-1}$, (c) $T$ = 1150 °C and $\dot{\varepsilon}$ = 0.001 s$^{-1}$, and (d) $T$ = 1150 °C and $\dot{\varepsilon}$ = 0.1 s$^{-1}$. Four true strain values are selected being 0, -0.24, -0.41, and -0.65. The results are taken from the set of simulations with a friction coefficient of 0.3. White and black arrows point out the failed elements that are deleted from the model.

In the GTN model, materials failure is modeled through void volume fraction, which is evolving to strain. **Figs. 9-10** respectively plot void volume fraction due to growth (VVFG) and void volume fraction due to nucleation (VVFN) as a function of strain from the element that first fails (i.e., the element that has the maximal evolution of VVFG and VVFN among all the elements). It can be seen that the VVFG decreases to a value slightly smaller than zero upon the increase of the strain, which is due to the closure of the initial voids under the compressive load. While VVFN keeps zero at the initial deformation stage. Then, the VVFG and VVFN vary very slightly and plateau with the increase of strain, respectively, and both suddenly rise very quickly till plateau starting from a critical strain. This critical strain varies among temperatures and strain rates, which is the failure strain $\varepsilon_f$ and its value is the same as the one extracted based on **Fig. 8**. In addition, based on **Figs. 9-10**, the magnitude of the failure strain $\varepsilon_f$ increases with the increase of temperature and with the decrease of strain rate, with temperature playing a more important role than strain rate, which is consistent with the aforementioned analysis.

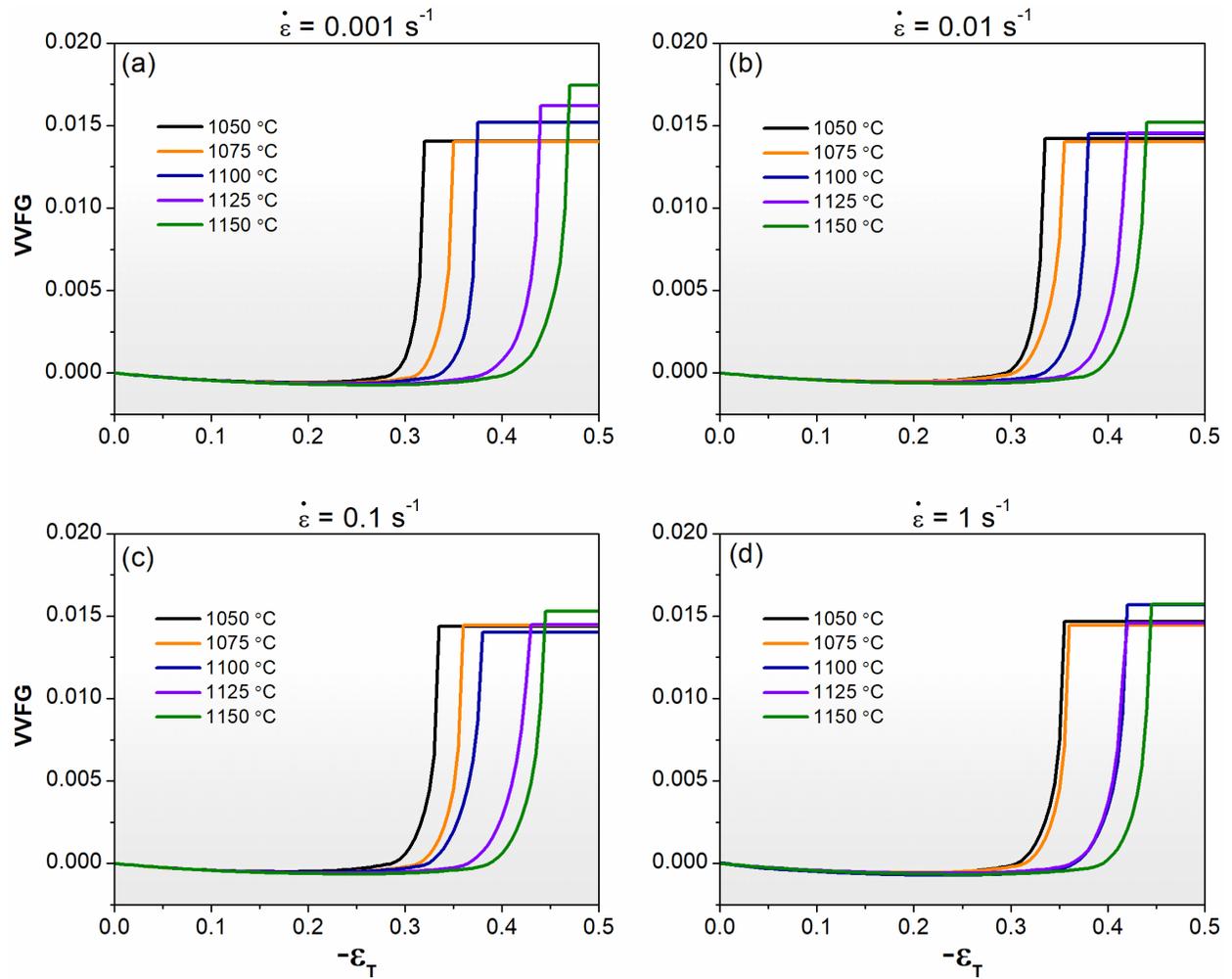

**Figure 9**. The evolution of void volume fraction due to growth (VVFG) for the first failed element (i.e., the element with the maximal VVFG) among all the elements in the FEA simulation as a function of external strain for different temperatures under strain rate in (a) 0.001 s$^{-1}$, (b) 0.01 s$^{-1}$, (c) 0.1 s$^{-1}$, and (d) 1 s$^{-1}$. The friction coefficient used is 0.3.

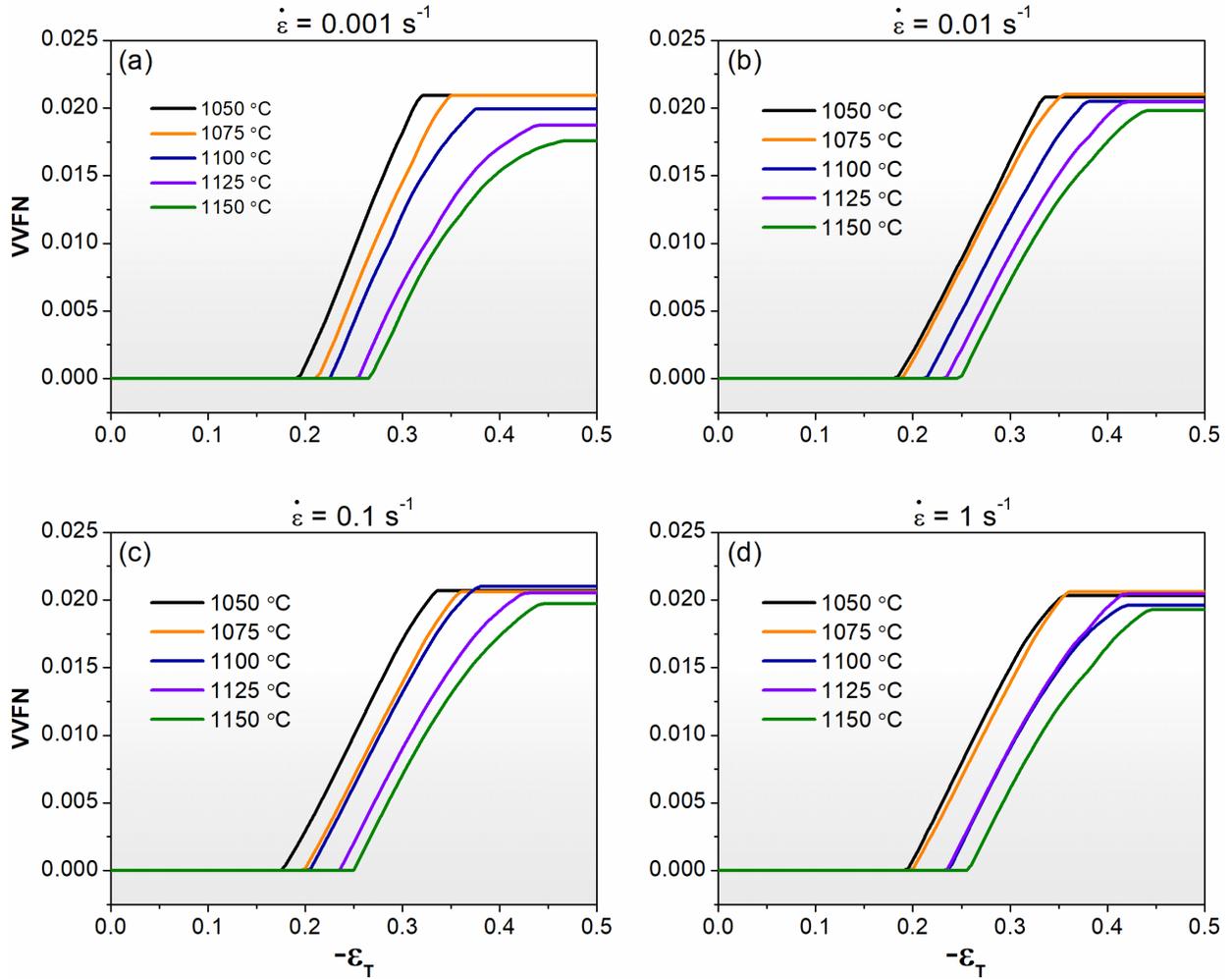

**Figure 10**. The evolution of void volume fraction due to nucleation (VVFN) for the first failed element (i.e., the element with the maximal VVFN) among all the elements in the FEA simulation as a function of external strain for different temperatures under strain rate in (a) 0.001 s$^{-1}$, (b) 0.01 s$^{-1}$, (c) 0.1 s$^{-1}$, and (d) 1 s$^{-1}$. The friction coefficient used is 0.3.

To quantitatively examine the influence of strain rate and temperature on the failure strain $\varepsilon_f$, the average and standard deviation of $\varepsilon_f$ calculated from different temperatures as a function of strain rate and calculated from different strain rates as a function of temperature are plotted in **Fig. 11a-b** and **Fig. 11c-d**, respectively. In addition, the effects of friction coefficients are also quantified as shown in **Fig. 11**. From **Fig. 11a-b**, it can be seen that the average $\varepsilon_f$ increases very slightly with the increase of strain, remaining nearly insensitive to strain rate with the slope of the linearly fitted curve being 0.0025 and 0.0020 for friction coefficients of 0.3 and 1.0, respectively. While for the same strain rate, the standard deviation is significant, indicating that the $\varepsilon_f$ varies greatly to the deformation temperature. The strong dependence of $\varepsilon_f$ on the temperature can be further revealed in **Fig. 11c-d**, where $\varepsilon_f$ increases in a third-order polynomial with temperature ($-\varepsilon_f = -0.73T^3 + 6.60 \times 10^{-4}T^2 - 1.98 \times 10^{-7}T + 269.03$ and $-0.79T^3 + 7.14 \times 10^{-4}T^2 - 2.15 \times 10^{-7}T + 288.97$ for friction coefficient being 0.3 and 1.0, respectively) and the standard deviation for a specific temperature is relatively very small. Therefore, it can be concluded that the deformation temperature plays a dominating role in determining the failure strain over the strain rate. The higher the deformation temperature, the larger the failure strain, which is preferred during hot working. Besides, the increasing trend of the $\varepsilon_f$ with the increase of strain rate and temperature agrees with He et al.'s [7] study, albeit that $\varepsilon_f$ in the present study is less sensitive to the strain rate. Moreover, the influence of the friction coefficient on the failure strain is found to be negligible.

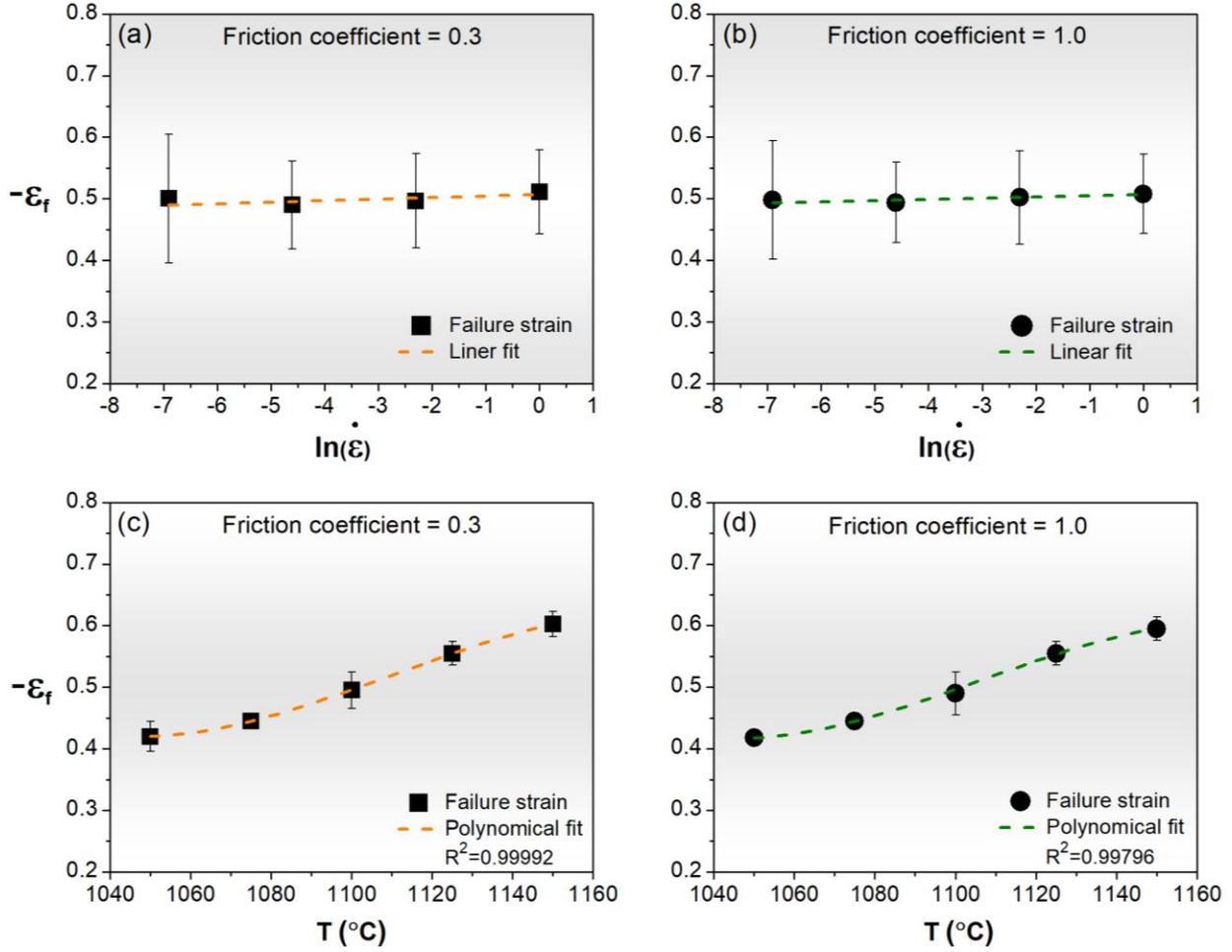

**Figure 11**. The average and standard deviation of the failure strain $-\varepsilon_f$ calculated from the range of temperatures as a function of strain rate in (a) and (b), and calculated from the range of strain rates as a function of temperature in (c) and (d). Friction coefficients used are 0.3 for (a) and (c) and 1.0 for (b) and (d).

With the failure strain numerically determined, processing maps with failure domains for different external strains $-\varepsilon_T$ can be determined, as shown in **Fig. 12**.

Firstly, for the flow instability $\xi(\dot{\varepsilon})$ (see the yellow shaded regions in **Fig. 12a-b**), it can be seen that it largely occurs at $-\varepsilon_T = 0.1$ either when the deformation temperature is lower than 1060 °C and the strain rate is moderate from 0.01 to 0.1 s$^{-1}$, or when the deformation temperature and strain rate are in the range from 1110 to 1150 °C and from 0.1 to 1 s$^{-1}$, respectively. Then, as shown in **Fig. 12b**, when $-\varepsilon_T$ increases to 0.2, the regions of the flow instability diminish to a localized region, where the temperature is around 1130 °C and the strain rate is from 0.3 to 1.0 s$^{-1}$. Further with the increase of the compressive strain, flow instability vanishes. Therefore, the deformation window of the alloy in consideration of avoiding flow instability should be at least $-\varepsilon_T > 0.2$.

Secondly, judging from the value of the black labels in **Fig. 12**, overall, the efficiency of power dissipation $\eta$ is increasing with the increase of the external compressive strain $-\varepsilon_T$ from 0.1 to 0.5, and decreases a bit at $-\varepsilon_T = 0.6$. This observation is also true for the evolution of the cyan-shaded regions to $-\varepsilon_T$, where the area of the regions undergoes sharp growth for $-\varepsilon_T$ from 0.1 to 0.3, slow growth for $-\varepsilon_T$ from 0.3 to 0.5, and a slight decrease for $-\varepsilon_T$ from 0.5 to 0.6. The strain rate that permits $\eta \geq 0.7$ is very low in the range from 0.001 to 0.003 s$^{-1}$, while the deformation temperature can be in a wide range from 1050 to 1150 °C except that the temperature should be lower than 1120 °C for $-\varepsilon_T = 0.1$. As larger $\eta$ represents better hot deformation properties of materials, the above results indicate that the hot deformation window for the alloy is comprised of $-\varepsilon_T$ from 0.3 to 0.6, very low strain rate in the range from 0.001 to 0.003 s$^{-1}$, and temperature in the range from 1050 to 1150 °C.

Next, it is observed that the failure first occurs at $-\varepsilon_T = 0.4$ for temperature lower than 1060 °C and strain rate as low as 0.001 s$^{-1}$. This phenomenon can be attributed to the high degree of work hardening at low temperatures and a long time of intense stress imposed on the samples during low strain rates to permit sufficient time for damage initiation and propagation. Then, at $-\varepsilon_T = 0.5$, the samples will mostly fail once the temperature is lower than 1110 °C regardless of the strain rate. This hints that the deformation temperature plays a dominating role over the strain rate in terms of failure, which echoes the aforementioned analysis. Moreover, the value in pink labels, i.e., the failure strain $\varepsilon_f$, increases with the increase of strain rate. This trend means the sample will occur failure earlier under a lower strain rate, in agreement with the above analysis for failure in **Fig. 12d**. This finding is also in line with the experimental results by He et al. [7], where the sample failed earlier under a lower strain rate for the same deformation temperature. Furthermore, at $-\varepsilon_T = 0.6$, the sample will fail under all the deformation conditions considered, except a small deformation window with a temperature higher than 1140 °C and strain rate as low as 0.0015 s$^{-1}$. Thus, to avoid failure of the sample, the deformation window in consideration of failure should be an external compressive strain $-\varepsilon_T < 0.6$ and temperature higher than 1110 °C.

Finally, the deformation window in light of the flow instability, efficiency of power dissipation, and failure can be defined for the Ni-based superalloy. The deformation can be around 0.5 in terms of compressive true strain that is corresponding to a 40% nominal compression of the sample to avoid failure, but no less than 0.2 to avoid flow instability. The deformation temperature should be in the range of 1110 ~ 1150 °C to avoid failure. The strain rate should be in the range from 0.001 to 0.002 s$^{-1}$ to guarantee relatively large power dissipation.

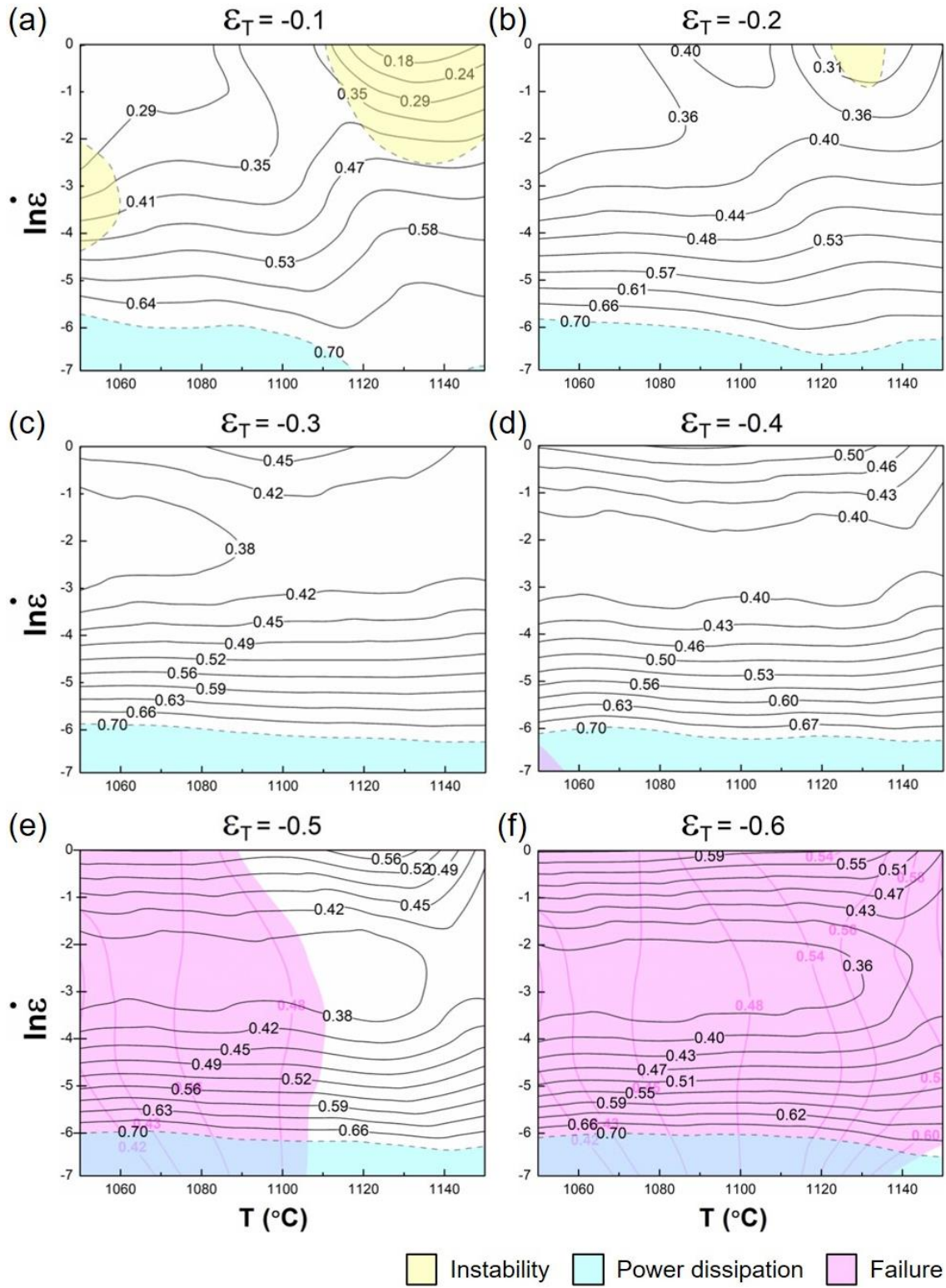

**Figure 12**. Processing maps comprised of power dissipation, instability, and failure maps for external true strain $-\varepsilon_T$ in (a) 0.1, (b) 0.2, (c) 0.3, (d) 0.4, (e) 0.5, and (f) 0.6. The black and red

lines and labels are for the power dissipation and failure maps, respectively. The yellow, cyan and pink shaded areas indicate regions of instability, power dissipation equal to or greater than 0.7, and failure, respectively. The data shown is from the set of simulations with a friction coefficient of 0.3.

## 4 Conclusions

In summary, we employed isothermal hot compression experiments and FEA to study the deformation and failure of a P/M Ni-based superalloy. Compression of the sample under different temperatures and strain rates was performed till a nominal strain of 50%. The GTN model with failure criterion was employed in the FEA to model the nucleation and growth of the voids in the superalloy. Experimentally, it is found that the stress-strain curves of the superalloy exhibit typical dynamic recovery and dynamic recrystallization rheological characteristics, and the failure morphologies are surface stripes along the loading axis of the sample. Next, with FEA simulations, it is demonstrated that the maximum tensile stress concentrates at the center of the outer bulging surfaces, which causes failure to initiate and propagate, being consistent with the experimental observations. In addition, the variation of the friction coefficient between 0.3 and 1.0 doesn't evidently affect the failure threshold. Then, the critical strain that initiates the failure is determined by FEA and it is further revealed that temperature dominates over strain rate in determining the critical failure strain. Specifically, the critical failure strain remains nearly insensitive to strain rate and increases in a third-order polynomial with the increase in temperature. Furthermore, processing maps constituted by the flow instability map, power dissipation map, and failure map are constructed to determine the optimized window for hot deformation of the superalloy. A 40% nominal compression of the superalloy under temperatures in the range of 1110 ~ 1150 °C and strain rates in the range from 0.001 to 0.002 $s^{-1}$ are formulated. The present study establishes a modeling framework for predicting the failure threshold for hot deformation of P/M superalloys and would be valuable in optimizing the hot working window of such alloys.

**Declaration of Competing Interest**

The authors declare that they have no known competing financial interests or personal relationships that could have appeared to influence the work reported in this paper.

**Acknowledgments**

Fanchao Meng acknowledges financial support from the National Key Research and Development Program of China (2021YFB3700403).